\documentclass[nameyear,MSNbibl,dvips]{arxstspdf}
\usepackage{graphicx}
\usepackage{flushend}
\usepackage{stfloats}

\volume{26}
\issue{1}
\pubyear{2011}
\firstpage{53}
\lastpage{56}
\doi{10.1214/11-STS345B}
\referstodoi{10.1214/10-STS345}

\begin{document}
\begin{frontmatter}

\vspace*{12pt}
\title{Discussion of ``Feature Matching in Time Series Modeling'' by Y. Xia and H. Tong}
\runtitle{Discussion}
\pdftitle{Discussion of Feature Matching in Time Series Modeling by Y. Xia and H. Tong}

\begin{aug}
\author{\fnms{Kung-Sik} \snm{Chan}\ead[label=e1]{kung-sik-chan@uiowa.edu}}
\and
\author{\fnms{Ruey S.} \snm{Tsay}\corref{}\ead[label=e2,text={ruey.tsay@ chicagobooth.edu}]{ruey.tsay@chicagobooth.edu}}
\runauthor{K.-S. Chan and R. S. Tsay}

\affiliation{University of Iowa and University of Chicago}

\address{Kung-Sik Chan is Professor, Department of Statistics and Actuarial Science,
University of Iowa, Iowa City, Iowa 52242, USA \printead{e1}. Ruey S.
Tsay is H. G. B. Alexander Professor, Booth School of Business, University of Chicago,
5807 South Woodlawn Avenue, Chicago, Illinois 60637,  USA
\printead{e2}.}

\end{aug}



\end{frontmatter}

We thank Xia and Tong for their stimulating article on time series
modeling. Their emphasis on estimation rather than model specification
is interesting. It brings new light to statistical applications in
general and to time series analysis in particular. The use of maximum
likelihood or least squares method is so common, especially with the
widely available statistical software packages, that one tends to
overlook its limitations and shortcomings.

There is hardly any statistical method or procedure that is truly
``one-size-fits-all'' in real applications. We welcome Xia and Tong's
contributions as they argue so convincingly that feature matching often
fares better  in time series modeling. On the other hand, we'd like to
point out some issues that deserve a careful study.

\section{Higher Order Properties}

The conditional mean function generally provides a~good description of
the cyclical behavior of the underlying process, and the catch-all
approach can be effectively implemented by estimating the model that
matches the multi-step conditional means to the data, as eminently
illustrated by the authors. Here, we want to point out the natural
extension of estimating a model by matching multi-step  conditional
higher moments to the data. For example, in financial time series
analysis, it is pertinent to model the dynamics of the conditional
variances. Consider the simple case that a time series of returns,
$\{r_t\}$, follows a generalized autoregressive conditional
heteroscedastic model of order (1, 1) or simply a $\operatorname{GARCH}(1,1)$ model:
\begin{eqnarray*}
r_t&=&\sigma_{t|t-1} \varepsilon_t, \\
\sigma_{t|t-1}^2&=&\omega+ \alpha r_{t-1}^2 +\beta \sigma_{t-1|t-2}^2,
\end{eqnarray*}
where $\omega > 0, \alpha \ge 0, \beta\ge 0$, $1 > \alpha+\beta> 0$ are
parameters, $\{\varepsilon_t\}$ are independent and identically
distributed (i.i.d.) random variables with zero mean and unit variance,
and $\varepsilon_t$ is independent of past one-step-ahead conditional
variances $\sigma^2_{s|s-1}, s\le t$. Estimation of the $\operatorname{GARCH}$ model can
be done by maximizing the Gaussian likelihood of the data, which
essentially matches the conditional variances with the squared returns.

A natural generalization of the catch-all method is to estimate a $\operatorname{GARCH}$
model that matches the $k$-step-ahead conditional variance to the $k$th
ahead data, for $k=1,2,\ldots,m$ with a fixed $m$, by minimizing some
weighted measure of dissimilarity of the multi-step conditional
variances to future squared returns. Various dissimilarity measures may
be used. Here, we illustrate the usefulness of this idea by adopting
the negative twice Gaussian log-likelihood as the dissimilarity
measure, that is, estimating\break a~$\operatorname{GARCH}$ model by minimizing
\[
S(\omega, \alpha, \beta)= \sum_{t=1}^{n-m} \sum_{\ell=1}^m
w_\ell \{ { r_{t+\ell}^2 }/{ \sigma^2_{t+\ell|t} }+\log(
\sigma^2_{t+\ell|t})\},
\]
where $\{w_\ell\}$ is a set of fixed weights and $\sigma^2_{t+\ell|t}$
is the conditional variance of $r_{t+\ell}$ given information available
at time $t$. When $m=1$, the new method reduces to the Gaussian
likelihood method. On the other hand, under the assumption that the
true model is a~$\operatorname{GARCH}$ model and for a fixed $m>1$, the estimator is
expected to be consistent and asymptotically normal, with details of the investigation
to be reported elsewhere. However, if the
$\operatorname{GARCH}$ model does not contain the true model, as likely is the case in
practice, the (generalized) catch-all method with $m>1$  may provide
new information for estimating a $\operatorname{GARCH}$ model that better matches the
observed volatility clustering pattern.

\begin{figure*}

\includegraphics{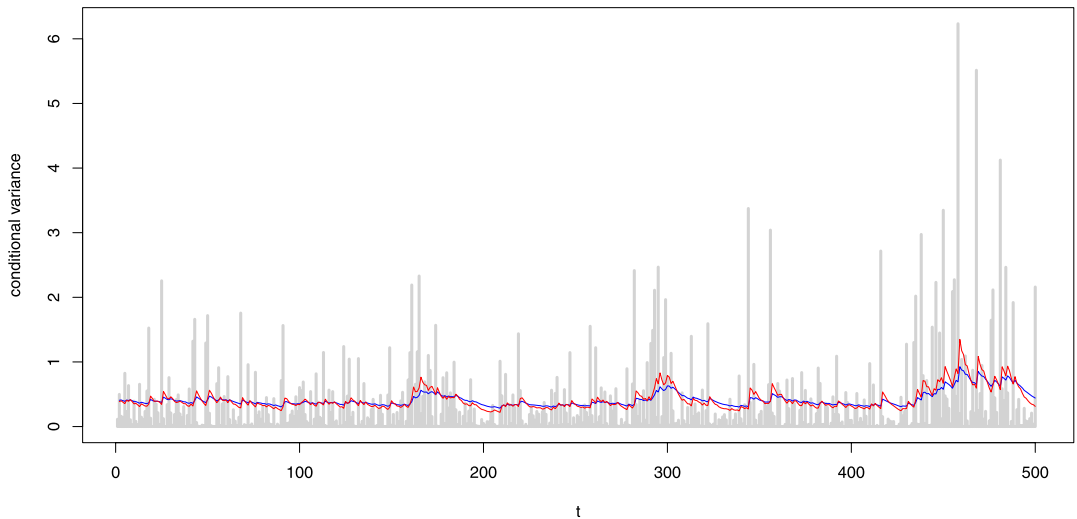}

        \caption{The red line connects the  one-step-ahead conditional variance,
        $\hat{\sigma}^2_{t|t-1}$, from the $\operatorname{GARCH}(1,1)$ model fitted by the
        catch-all method with $m=30$ and equal weights,
        whereas their counterparts from the model fitted by the
        Gaussian likelihood method are connected by  the blue line.
        The light gray bars are the squared daily returns of the CREF series.}
        \label{fig:1}
         \end{figure*}

We tried this approach by fitting a $\operatorname{GARCH}(1,1)$ model to the daily
returns of a unit of the CREF stock fund over the period from August
26, 2004 to August 15, 2006; this series was analyzed by Cryer and Chan
[(\citeyear{CryCha08}), Chapter 12], and they identified the series as a $\operatorname{GARCH}(1,1)$
process. Gaussian likelihood estimation yields $\hat{\omega}=0.0164,
\hat{\alpha}=0.0439, \hat{\beta}=0.917$. On the other hand, the
catch-all method, with $m=30$ and the weights $w_\ell\propto 1$, results
in the estima\-tes: $\hat{\omega}=0.0261, \hat{\alpha}=0.102$ and
$\hat{\beta}=0.836$. Figu\-re~\ref{fig:1} contrasts the fitted values,
$\hat{\sigma}^2_{t|t-1}$, based on the two fitted $\operatorname{GARCH}$ models with the
squared returns superimposed as light gray bars. The figure shows that,
as compared to the model estimated by the Gaussian likelihood method,
the $\operatorname{GARCH}$ model fitted by the catch-all method appears to track the
squared~re\-turns more closely over the volatile period and transit into
the ensuing quiet period at a faster rate commensurate with the data.
It seems that by~\mbox{requiring} the model to track multi-step squared
returns, the method chooses a $\operatorname{GARCH}$ model that gives more weight to the
current squared return to the future evolution of conditional
variances. Indeed, as $m$ increases from 1 to $30$, the ARCH
coefficient~\mbox{estimate}~$\hat{\alpha}$ increases from 0.0439 to 0.102,
whereas the $\operatorname{GARCH}$ coefficient $\hat{\beta}$ decreases from 0.917 to
0.836. These~systematic parametric changes signify that~the true~mo\-del
is unlikely a $\operatorname{GARCH}(1,1)$ model.\vadjust{\goodbreak} It also matches nicely with the
increasing kurtosis of the data; see Tsay [(\citeyear{Tsa05}), Chapter~3] for a
discussion~on contribution of $\alpha$ to kurtosis of the return $r_t$.
This example illustrates the usefulness of the generalization of the
catch-all method by matching higher moments, and also the potential
usefulness of developing a test for model misspecification based on
the divergence of the catch-all estimates with increasing~$m$.

\section{Information Content}

As statisticians, we love data. However, data have their limitation.
Available data may not be informative to conduct any meaningful feature
matching. When the information content pertaining to the selected
feature is low, feature matching as an estimation method is likely to
fail. As an example, assessment of financial risk focuses on the tail
behavior of the loss over time. The relevant feature here is the upper
quantiles of the loss distribution. Since big losses are rare,
available data, no matter how big the sample size is, are not
informative about the extreme quantiles. The uncertainty in any
matching would be high.

As another example, consider the monthly time series of global
temperature anomalies from 1880 to 2010. The data are available from
many sources on the web, for example, the websites
\href{http://data.giss.nasa.gov/gistemp/}{http://data.giss.}
\href{http://data.giss.nasa.gov/gistemp/}{nasa.gov/gistemp/}
of the Goddard Institute for Spa\-ce
Studies (GISS), National Aeronautics and Space Administration
(NASA)
and \href{http://www.ncdc.noaa.gov/cmb-faq/anomalies.html}{http://www.ncdc.noaa.}
\href{http://www.ncdc.noaa.gov/cmb-faq/anomalies.html}{gov/cmb-faq/anomalies.html}\vadjust{\goodbreak}
of the National Clima\-tic
Data Center, National Oceanic
and Atmospheric Administration (NOAA).
These time series are of interest because of the concerns about global
warming. The key feature to match then is the underlying trend of the
global temperature. To handle the time trend, two approaches are
commonly used in the time series literature. The first approach is
called the difference-stationarity in which the time series~is
differenced to obtain stationarity. The $\operatorname{ARIMA}$ models of Box and Jenkins
are examples of this \mbox{approach}. The second approach is called the
trend-stationarity in which one imposes a linear time trend for the~data.
The deviation from the time trend becomes a~statio\-nary series.
Even though we have 130 years of~data,~it remains hard for the data to
distinguish a~\mbox{trend-sta-}\break tionary model from a difference-stationary one.
On the other hand, the long-term forecasts of a~differen\-ce-stationary
model differ dramatically from those of a trend-stationary model. The
eventual forecasting function of an $\operatorname{ARIMA}$ model without
constant is a~horizontal line with standard error approaching infinity whereas that
of a~trend-stationary model is a straight line going to positive or
negative infinity with a finite standard error. See Tsay (\citeyear{Tsa12}) for
further analysis of the data and model comparison.\looseness=-1

Here, we explore whether feature matching maycast new light on the
preceding problem. For simplicity, we consider the annual global
temperature anomalies. Model identification suggests two plausible
models, namely,  $\operatorname{ARIMA}(1,1,1)$ model and linear trend plus
$\operatorname{ARMA}(1,1)$
model. We use the plus convention for the $\operatorname{MA}$ coefficients, that is,
the\break
$\operatorname{ARIMA}(1,1,1)$ model specifies $(1-\phi B)(1-B)Y_t= (1+\theta B)a_t$,
where $B$ is the backshift operator such that $BY_t = Y_{t-1}$ and
$a_t$ are i.i.d. with zero mean and variance $\sigma_a^2$. For
simplicity, these models are fitted by conditional least squares with
the residuals initialized as 0 at the first time point. We also fitted
the model with the generalized catch-all method that matches the
predictive distribution to the future $m$ values, in terms of  the
$k$-step-ahead predictive means and variances, $k=1,2,\ldots,m$. The
loss function is similar to that discussed in the previous section.
After profiling out the innovation variance, the generalized catch-all
method fits a model to the time series $\{Y_t \mid t=1,2,\ldots,T\}$ by
minimizing
\[
S(\theta)=\sum_t \sum_{\ell=1}^m
(Y_{t+\ell}-\hat{Y}_{t+\ell|t})^2\overline{\sigma^2}/\sigma^2_\ell,
\]
%
\begin{figure}

\includegraphics{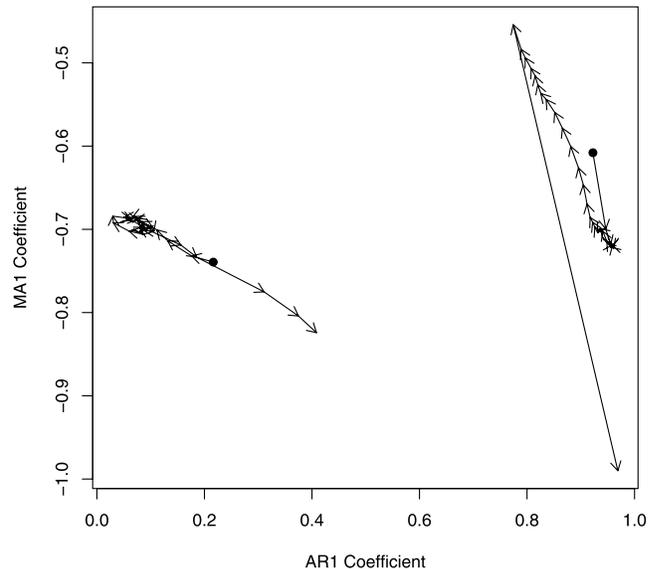}
        \caption{ The directed scatterplot on the right side shows the evolution of the $\operatorname{ARMA}$ coefficients
        for the linear trend plus $\operatorname{ARMA}(1,1)$ noise model fitted to
        the annual global temperature anomalies, with
        $m$ in the catch-all method increasing from 1 to 30, while
        the evolution of the $\operatorname{ARMA}$ coefficients for the $\operatorname{ARMA}(1,1,1)$ model is
        shown by the directed scatterplot on the left.
        The two solid circles show the estimates with $m=1$.}
\label{fig:2}
\end{figure}
where $\theta$ is the vector of all parameters  except the innovation
variance;\vadjust{\goodbreak} $\hat{Y}_{t+\ell|t}$ is the $\ell$-step predictive mean,
$\sigma^2_\ell$ is the $\ell$-step-ahead prediction variance,  that is, $ \sigma^2_\ell=\sigma_a^2
\sum_{j=0}^{\ell-1} \psi_j^2$ where $\psi$'s are the coefficients in
the linear $\operatorname{MA}$ representation of the model also known as the impulse
response coefficients; $\overline{\sigma^2}$ is the harmonic mean of
$\sigma^2_\ell, \ell=1,2,\ldots,m$. When $m=1$, the catch-all method
reduces to Gaussian~li\-kelihood estimation. But for $m>1$, the catch-all
method  attempts to match model prediction with~$m$ ``future'' values, in
terms of means and variances.~We implemented the catch-all method for
fitting the~glo\-bal temperature anomalies, with $m=1,2,\ldots,30$.~%
Fi\-gure~\ref{fig:2} plots the evolution of the catch-all estimates of
the $\operatorname{ARMA}$ coefficients for the two models. It is interesting to observe
that the catch-all estimates of the $\operatorname{ARIMA}(1,1,1)$ model quickly move
away from the Gaussian likelihood estimates and fluctuate stably around
an essentially $\operatorname{ARIMA}(0,1,1)$ model, that is, an exponential smoothing
model, for a~while,  before they drift to more extreme values. Thus,
the catch-all method suggests that for forecasting on a decadal scale,
an exponential smoothing model may be appropriate for the temperature
data. Similarly, the catch-all estimates of the linear trend plus
$\operatorname{ARMA}(1,1)$ noise model quickly move away from the Gaussian
likelihood
estimate, and fluctuate stably for a number of steps around an estimate
with its $\operatorname{MA}(1)$ coefficient comparable to that of the exponential
smoothing model. As the corresponding $\operatorname{AR}(1)$ estimates are quite close
to 1, the catch-all estimates of the trend plus noise model suggest
strong similarity between the two models over a forecast horizon on a
decadal scale. When $m$ approaches 30, the catch-all estimates of the
trend plus $\operatorname{ARMA}(1,1)$ noise model drift off on a flight that suddenly
ends into a trend plus uncorrelated noise model.

For long-term prediction, the global temperature  series is of limited
value. The preceding analysis highlights the conundrum that using a
trend-statio\-nary model, we simply force the data to support the trend;
for difference-stationary models, we basically end up using the
exponential smoothing model. This is a problem facing all estimation
methods, not just feature matching.

We consider these two examples not because we question the value of
feature matching in time series modeling; rather we like to point out
that care must be exercised in using feature matching. In other\break words,
feature matching may encounter the same problem as the traditional
estimation methods. They are statistical tools. It is the statisticians
who use the tools, not the tools that process the data. Which tool to
use in a given application is the choice of a~statistician. While we
welcome the addition of feature matching to the tool kits, we like to
emphasize that there are limitations in feature matching, too.

\section{Feature Versus Objective}

Model selection in data analysis depends critically on the objective of
data analysis. Likelihood estimation seeks parameters that give the
highest probability of the data under the entertained model. Feature
matching searches for parameters that best describe the features of
interest. The examples used in the article all have clearly defined
features such as cycles and, as expected, feature matching works well.
On the other hand, there are situations under which the objective of
data analysis does not match well with any specific feature. For
example, with the economy under pressure, unemployment is of major
importance to people of all walks. It is well known that unemployment rate
exhibits strong business cycles, which in sharp contrast with examples
used in the article do not have a fixed (or even an approximate)
period. It is then not clear which feature to match if one is
interested in understanding and forecasting the U.S. unemployment
rate.\vspace*{24pt}

\section*{Acknowledgments}

K.-S. Chan thanks the U.S. National Science Foundation (NSF-0934617) and
R. S. Tsay thanks the Booth School of Business, University of Chicago, for partial financial support.

\vspace*{12pt}


\begin{thebibliography}{3}

\bibitem[\protect\citeauthoryear{Cryer and Chan}{2008}]{CryCha08}
\begin{bbook}[auto:STB|2011-03-03|12:04:44]
\bauthor{\bsnm{Cryer},~\bfnm{J.~D.}\binits{J.~D.}} \AND
  \bauthor{\bsnm{Chan},~\bfnm{K.~S.}\binits{K.~S.}}
(\byear{2008}).
\btitle{Time Series Analysis: With Applications in R},
\bedition{2nd ed.}
\bpublisher{Springer}, \baddress{New York}.
\end{bbook}
\endbibitem

\bibitem[\protect\citeauthoryear{Tsay}{2010}]{Tsa05}
\begin{bbook}[mr]
\bauthor{\bsnm{Tsay},~\bfnm{Ruey~S.}\binits{R.~S.}}
(\byear{2010}).
\btitle{Analysis of Financial Time Series},
\bedition{3rd ed.}
\bpublisher{Wiley}, \baddress{Hoboken, NJ}.
\end{bbook}
\endbibitem

\bibitem[\protect\citeauthoryear{Tsay}{2012}]{Tsa12}
\begin{bbook}[auto:STB|2011-03-03|12:04:44]
\bauthor{\bsnm{Tsay},~\bfnm{R.~S.}\binits{R.~S.}}
(\byear{2012}).
\btitle{An Introduction to Financial Data Ana\-lysis}.
\bpublisher{Wiley}, \baddress{Hoboken, NJ}.
\end{bbook}
\endbibitem

\end{thebibliography}
\end{document}